\documentclass[12pt,pra,aps,amssymb,amsfonts,amsmath,tightenlines]{revtex4}

\usepackage[dvips]{graphicx}
\usepackage{dsfont, color,txfonts} 

\newcommand{\re}{\mbox{$\rm e$}}
\newcommand{\rd}{\mbox{$\rm d$}}
\newcommand{\half}{\mbox{$\textstyle \frac{1}{2}$}}

\begin{document}

\title{Modelling election dynamics and the impact of disinformation}
\author{Dorje C. Brody}

\affiliation{
Department of Mathematics, University of Surrey, Guildford GU2 7XH, UK 
}

\date{\today}

\begin{abstract}
Complex dynamical systems driven by the unravelling of information can be modelled effectively by treating the underlying flow of information as the model input. Complicated dynamical behaviour of the system is then derived as an output. Such an information-based approach is in sharp contrast to the conventional mathematical modelling of information-driven systems whereby one attempts to come up with essentially \textit{ad hoc} models for the outputs. Here, dynamics of electoral competition is modelled by the specification of the flow of information relevant to election. The seemingly random evolution of the election poll statistics are then derived as model outputs, which in turn are used to study election prediction, impact of disinformation, and the optimal strategy for information management in an election campaign. 
\end{abstract}

\maketitle


\section{Introduction}

Modelling and understanding of democratic processes such as election or 
referendum have become increasingly important in recent years, in light of 
the potential threat to democracy posed by the large-scale dissemination of 
disinformation on the internet and in other media (cf. \cite{WD,DCMSC}). 
In particular, in response 
to the widely publicised circulation of disinformation during the 2016 US 
presidential election and the `Brexit' referendum in the UK on the membership 
of the European Union, a lot of research efforts have been devoted towards 
the detection, prevention, and retrospective analysis of false information 
circulated on the internet---the so-called \textit{fact checking}---for which there 
is already a substantial body of literature 
\cite{Allcott,Amador,bovet,CRC,KH,Shu,Yang}. Whilst important, fact 
checking alone, however, is insufficient to counter impacts of disinformation; 
what is equally important is the development of \textit{dynamical} modelling 
frameworks for democratic processes, and how dissemination of disinformation 
might affect them, with a view towards scenario analysis, impact studies, and 
strategic planning. 

In response to this demand we have recently introduced an information-based 
framework for modelling the dynamics of opinion-poll statistics in elections and 
referendums \cite{BM}. The idea that imperfect information about candidates' 
positions on issues, or about their competency and integrity, must play a 
significant role in modelling election has been argued previously, for instance 
in \cite{JH}, where game 
theory under incomplete information is applied to model electoral competition. 
In fact, various authors have explored the role of information in election. To 
name a few, in \cite{MO} conditions for the existence of an equilibrium state are 
established when there are informed and uninformed voters. The aggregation of 
information in election with imperfect information, and conditions for establishing 
equilibrium, are investigated in \cite{FP}. There are also empirical studies to show 
how voting pattern vary in accordance with how informed the electorate are 
\cite{FM}. 
In view of the prevalence of social 
media usage and the advances in mobile technology, where information can 
instantly be transmitted far and wide, the role of information 
in election modelling is now becoming acutely more important. With this in mind, 
here we explore further details of the election modelling by extending the work 
presented in \cite{BM}. 

What distinguishes the information-based approach of \cite{BM} from the 
previous work is that a dynamical model for the flow of information is specified 
as a starting point. How the flow of information would affect the dynamical 
evolution of the opinion poll is then \textit{derived} as an output, rather than 
being modelled. This is achieved by 
borrowing mathematical techniques from signal processing 
(cf. \cite{Wiener,kai}) of converting noisy 
information into the best estimates of the quantities of interests. Because the 
transformation of the underlying information process into the output is highly 
nonlinear, even with a relatively simplified model choice for the flow of information 
it is nevertheless possible to describe dynamical behaviours of complex 
systems in such a way that it is consistent with our intuition, as we shall 
demonstrate through various analysis and examples. 
(That this is the case has been demonstrated in financial modelling for the 
price processes of various assets, including credit-risky bonds \cite{BHM}, 
reinsurance contracts \cite{BHM2}, or crude oil \cite{BHY}.) When it comes to 
modelling electoral competition, in particular, we emphasise that our scheme 
offers, to our best knowledge, the first fully-dynamical framework that 
captures the impact of information revelation in a noisy environment; 
whereas previously proposed dynamical models are either 
deterministic or else are not dependent on information revelation 
(see, e.g., \cite{xxx,yyy,Lloyd}). This, 
in turn, allows us to work out probabilities of the occurrences of future 
events explicitly, as we shall demonstrate, and these results are 
indispensable  for scenario analysis, impact studies, and strategic planning. 

The paper is organised as follows. We begin in Section~\ref{sec:2} with a brief 
overview of the two approaches for modelling electoral competition proposed in 
\cite{BM}; one called structural approach and the other called reduced-form 
approach. Our focus in the present paper will be on the latter approach, which 
is developed further in Section~\ref{sec:3} with the analysis of identifying the 
arrival of new information in a noisy communication channel. The results are 
then applied in Section~\ref{sec:4} to derive the formula for the \textit{a priori} 
probability of a given candidate winning the election. The dynamical evolution 
of the winning probability is then worked out in Section~\ref{sec:5}. In 
Section~\ref{sec:6} we examine the impact of disinformation on the winning 
probability. In particular, we identify the optimal strategy for disinformation so 
as to maximally impact the winning probability, in the case of a very simple 
model for fake news. More generally, the management of information in an 
electoral competition is examined in Section~\ref{sec:7}, where we analyse 
how the winning probability is affected as we vary the information flow rate in 
a time-dependent setup. The sensitivity of the winning probability against the 
information flow rate parameter is analysed in Section~\ref{sec:8} by 
borrowing techniques from information geometry and working out the Fisher 
information. The result will be useful for the purpose of cost-benefit analysis 
in an advertisement campaign. 
We conclude in Section~\ref{sec:9} with a discussion on how the 
techniques developed in this paper can be applied, \textit{mutatis mutandis}, 
within the structural approach, to help election campaign in a realistic setup, 
and more generally to manage information in advertisement. We remark that 
the purpose of this paper is not in developing new mathematics, but rather to 
develop a novel information-based approach to model the impact of 
advertisement, with a focus on electoral competition. Indeed, most of the 
mathematical manipulations are elementary, and can be found, for instance, 
in \cite{BH,BHM}. With this declaration we shall omit repetitive citation to 
these papers when similar calculations are performed in different contexts.

\section{Information-based modelling frameworks for election}
\label{sec:2}

In \cite{BM} we introduced two closely-related approaches to model election 
dynamics: one is called an election-microstructure approach, or a 
\textit{structural approach}, which encapsulates 
structural details in a voting scenario; and the other is called a representative 
voter approach, or a \textit{reduced-form} approach, which captures qualitative 
features of election dynamics without the specification of structural details. Let 
us begin by briefly describing these approaches. 

In the structural approach, one considers a set of issues that are of concern to 
the electorates. These may include, for instance, the candidates' positions on 
social welfare, immigration, abortion, climate policy, gun control, healthcare, and 
so on. The $l$-th candidate's position on the $k$-th issue will then be represented 
by $X^l_k$, whose value is not necessarily apparent to the voters. The 
uncertainties for the factor $X^l_k$ thus make it a \textit{random variable} on a 
suitably defined probability space equipped with the `real-world' probability measure 
${\mathbb P}$. The idea is that, for instance, if the $k$-th 
factor were concerned with, say, climate policy, then in a simple model setup 
$X^l_k=0$ would represent the 
situation in which candidate $l$ is against implementing policies to counter 
climate change, while $X^l_k=1$ would represent the situation in which 
candidate $l$ is for implementing such policies. Not all factors need to be binary, 
of course, but at any rate the voters are not certain about the positions of the 
candidates on these issues, if they were elected. 

The voters, however, possess partial information concerning the values of these 
factors, and as they learn more about the candidate, or perhaps about the political 
party to which the candidate belongs, their best estimates for 
these factors will change in time. The voters also have their preferences: what is 
a desirable position of a candidate on a given issue to one voter may well be 
undesirable to another voter. Let us denote by $w^k_n$ the preference weight of 
the $n$-th voter for the $k$-th factor, which may be positive or negative, depending 
on the voter's position on that issue. Then writing ${\mathbb E}_t[X^l_k]$ for the 
expectation under the probability measure ${\mathbb P}$ conditional on the 
information available at time $t$, the \textit{score} at time $t$ assigned to the 
$l$-th candidate by the $n$-th voter, under the assumption of a linear scoring 
rule, is given by the sum 
\begin{eqnarray} 
S^l_n(t) = \sum_{k} w^k_n\, {\mathbb E}_t[X^l_k]\, . 
\end{eqnarray} 
In particular, the voter will select the candidate with the highest score at time 
$T\geq t$ when the election takes place. Thus by modelling the flow of 
information associated with each of the factors, we arrive at a rather detailed 
dynamical model for election, and this is the basis of the structural approach. 

In the present paper we shall be concerned primarily with the reduced-form 
approach and develop the theory in more detail. We consider an election in which 
there are $N$ candidates. In the reduced-form 
approach, the voters are in general not fully certain about which candidate they 
should be voting for, but they have their opinions based on (i) the information 
available to them about the candidates, or perhaps about the political party to 
which they belong, and (ii) the voter preferences. The diverse opinion held by 
the public can then be aggregated in the form of a probability distribution, 
representing the public preference likelihoods of the candidates. Thus we can 
think of an abstract random variable $X$ taking the value $x_i$ with the 
\textit{a priori} 
probability ${\mathbb P}(X=x_i)=p_i$ defined on a probability space, where 
$x_i$ represents the $i$-th candidate, and the \textit{a priori} probability $p_i$ 
represents today's opinion-poll statistics. 

Today's public opinion, however, changes over time in accordance with the 
unravelling of new information relevant to the election. Hence the \textit{a 
priori} probability will be updated accordingly, generating a shift in the 
opinion-poll statistics. To model this dynamics, let us assume (merely for 
simplicity) that reliable knowledge regarding which candidate represents the 
`best choice' increases linearly in time, at a constant rate $\sigma$. There 
is also a lot of rumours and speculations that obscure the reliable information 
in the form of noise. This uncertainty, or noise, will be modelled by a Brownian 
motion, denoted by $\{B_t\}$, which is assumed to be independent of $X$ 
because otherwise it cannot be viewed as representing pure noise. Hence, 
under these modelling assumptions the 
flow of information, which we denote by $\{\xi_t\}$, can be expressed in the form 
\begin{eqnarray}
\xi_t = \sigma X t + B_t . 
\label{eq:1} 
\end{eqnarray} 
For the voters, the quantity of interest is the actual value of $X$, but there are 
two unknowns, 
$X$ and $\{B_t\}$, and only one known, $\{\xi_t\}$. In this situation, a rational 
individual considers the probability that $X=x_i$ conditional on the information 
contained in the time series $\{\xi_s\}_{0\leq s\leq t}$ gathered up to time $t$. 

We proceed to determine the conditional probability 
${\mathbb P}(X=x_i|\{\xi_s\}_{0\leq s\leq t})$. We begin by remarking that the 
information process $\{\xi_t\}$ is Markovian. An intuitive way of seeing this is 
to observe that the increment $\rd\xi_t$ of $\{\xi_t\}$ is given by the sum of 
$\sigma X \rd t$ 
and $\rd B_t$, but the coefficient of $\rd t$ is constant in time, while the Brownian 
motion has independent increments, so the process $\{\xi_t\}$ of (\ref{eq:1}) 
does not carry any `memory'. Establishing the Markov property is equivalent to 
showing that $ {\mathbb P}(\xi_t\leq x | \xi_s,\xi_{s_1},\xi_{s_2},\ldots,\xi_{s_k}) 
= {\mathbb P}( \xi_t\leq x|\xi_s)$ for any collection of ordered times $t, s, s_1, 
s_2, \ldots,s_k$. However, the process $\{\xi_t\}$ conditional on $X=x_i$ is just 
a drifted Brownian motion, which clearly is Markovian, so we have 
\begin{eqnarray}
{\mathbb P}\left( \xi_t\leq x| \xi_s,\xi_{s_1}, 
\ldots,\xi_{s_k}\right)&=& \sum_i {\mathbb P}\left( \xi_t\leq x| X=x_i, 
\xi_s,\xi_{s_1}, \ldots,\xi_{s_k}\right) {\mathbb P}(X=x_i) 
\nonumber \\ &=&
\sum_i {\mathbb P}\left( \xi_t\leq x| X=x_i,\xi_s\right) {\mathbb P}(X=x_i) 
\nonumber \\ &=&
{\mathbb P}\left( \xi_t\leq x\Big| \xi_s\right).
\end{eqnarray}
In addition to the Markovian property, the random variable $X$ is a function 
of the time series $\{\xi_t\}$ because, with probability one we have 
\begin{eqnarray} 
\lim_{t\to\infty} \frac{\xi_t}{\sigma t} = X . 
\end{eqnarray} 
It follows that the conditional probability 
${\mathbb P}(X=x_i|\{\xi_s\}_{0\leq s\leq t})$ simplifies to 
${\mathbb P}(X=x_i|\xi_t)$. 

The logical step of converting the prior probabilities ${\mathbb P}(X=x_i)$ into 
the posterior probability ${\mathbb P}(X=x_i|\xi_t)$ is captured by the Bayes 
formula: 
\begin{eqnarray}
{\mathbb P}(X=x_i|\xi_t) &=&  \frac{{\mathbb P}(X=x_i)\rho
(\xi_t|X=x_i)}{\sum_{j} {\mathbb P} (X=x_j)
\rho(\xi_t|X=x_j)} . 
\label{eq:c2.33}
\end{eqnarray}
Here the conditional density function $\rho(\xi_t|X=x_i)$ for the random variable 
$\xi_t$ is defined by the relation
\begin{eqnarray}
{\mathbb P}\left(\xi_t\leq y|X=x_i\right)=\int_{-\infty}^y
\rho(\xi|X=x_i)\,\rd\xi, 
\label{eq:c2.34}
\end{eqnarray}
and is given by
\begin{eqnarray}
\rho(\xi|X=x_i)=\frac{1}{\sqrt{2\pi t}} \exp\left(-
\frac{(\xi-\sigma x_i t)^2}{2t}\right). 
\label{eq:c2.35}
\end{eqnarray}
This follows from the fact that, conditional on $X=x_i$, the random variable 
$\xi_t$ is normally distributed with mean $\sigma x_i t$ and variance $t$. 
We thus deduce that 
\begin{eqnarray}
{\mathbb P}(X=x_i|\xi_t) =\frac{p_i\exp\left( \sigma x_i \xi_t-
\frac{1}{2} \sigma^2 x_i^2 t\right)} {\sum_j p_j 
\exp\left( \sigma x_j \xi_t-\frac{1}{2} \sigma^2 x_j^2 t\right)} .
\label{eq:5}
\end{eqnarray}
Inferences based on the use of (\ref{eq:5}) are optimal in the sense that they 
minimise the uncertainty concerning the value of 
$X$, as measured by the variance or entropic measures subject to the information 
available. Thus the \textit{a posteriori} probabilities (\ref{eq:5}) determine 
the best estimate for the unknown variable $X$, in the sense of minimising 
the error. 

In the reduced-form approach it is the conditional probability (\ref{eq:5}), which is 
a nonlinear function of the model input $\{\xi_t\}$, that models the complicated 
dynamics of the opinion poll statistics. Our objective in this paper is to investigate 
various properties of the model, as well as to explore different ways in which the 
model can be exploited.

\section{Election dynamics and the arrival of new information} 
\label{sec:3}

We begin by considering the dynamical evolution of the conditional probability 
obtained in (\ref{eq:5}). For this purpose let us introduce a simpler notation by 
writing $\pi_{it}={\mathbb P}(X=x_i|\xi_t)$. Then an application of Ito's formula 
on (\ref{eq:5}) yields 
\begin{eqnarray} 
\rd\pi_{it} = \sigma( x_i-{\hat X}_t) \pi_{it} \left( \rd \xi_t - \sigma 
{\hat X}_t \rd t \right), 
\end{eqnarray}
where we have written 
\begin{eqnarray}
{\hat X}_t = \sum_{i} x_i {\mathbb P}(X=x_i|\xi_t) 
\label{eq:11} 
\end{eqnarray} 
for the conditional expectation of $X$ given $\xi_t$. 

Examining the dynamics of the conditional probability is of interest because 
it allows us to isolate the arrival of new information. We remark that the time 
series $\{\xi_t\}$ in itself contains new information as well as redundant 
information that has already been identified. (It is often the case that when 
one reads a newspaper article, for instance, only a fraction of what one reads 
is genuinely new---and this tendency is more pronounced with the web-generated 
news contents, which is becoming mainstream.) That the conditional 
expectation (\ref{eq:11}) gives the `best' 
estimate of $X$ is then reflected in the remarkable fact that at time $t$, the 
small increment in the electorate's belief, represented by $\rd\pi_{it}$, is induced 
only by the new information. In other words, the small increment $\rd W_t$ of the 
process $\{W_t\}$ defined by  
\begin{eqnarray} 
W_t = \xi_t - \sigma \int_0^t {\hat X}_u \rd u 
\label{eq:12} 
\end{eqnarray} 
represents the arrival of new information. 

The process $\{W_t\}$ defined in (\ref{eq:12}) is known as the \textit{innovations 
process} in the theory of signal processing \cite{kai}. It can be shown, in fact, that 
$\{W_t\}$ is a standard Brownian motion. For this purpose, we recall the L\'evy 
criterion for Brownian motion that if a process $\{W_t\}$ is a martingale, and if 
$(\rd W_t)^2=\rd t$, then $\{W_t\}$ is a Brownian motion. To show 
that $\{W_t\}$ is a martingale, we observe for $t\geq s$ that 
\begin{eqnarray}
{\mathbb E}_s\left[W_t\right]&=& {\mathbb E}_s[\xi_t]-\sigma \, 
{\mathbb E}_s\left[\int_0^t {\hat X}_u {\rd}u\right]
\nonumber \\ &=&  \sigma t {\hat X}_s + {\mathbb E}_s[ B_t] - 
\sigma \int_0^s {\hat X}_u \rd u - \sigma (t-s) {\hat X}_s ,
\label{eq:96}
\end{eqnarray}
but from the tower property of conditional expectation we have 
${\mathbb E}_s[B_t]={\mathbb E}_s[B_s]$, so writing 
${\hat X}_s={\mathbb E}_s[X]$ in (\ref{eq:96}) we find, on account of 
${\mathbb E}_s[ \sigma s X + B_s] = {\mathbb E}_s[ \xi_s]=\xi_s$, that  
\begin{eqnarray}
{\mathbb E}_s\left[W_t\right] = {\mathbb E}_s[ \sigma s X + B_s] - 
\sigma \int_0^s {\hat X}_u \rd u = W_s \, ,
\end{eqnarray}
establishing the martingale condition. Then from $(\rd \xi_t)^2=\rd t$ we find at 
once that $(\rd W_t)^2=\rd t$, from which it follows that $\{W_t\}$ is a Brownian 
motion under the measure ${\mathbb P}$ with respect to the information flow 
generated by $\{\xi_t\}$. 

With this observation at hand, we infer from (\ref{eq:12}) that the information 
process $\{\xi_t\}$ in the `real-world' probability measure ${\mathbb P}$ is a 
drifted Brownian motion, which means that there exists a fictitious probability 
measure ${\mathbb Q}$ under which the information process $\{\xi_t\}$ is 
itself a Brownian motion. The details of this measure change becomes handy 
in the analysis below for the probability of a given candidate winning the election. 
To clarify the relation between the two measures ${\mathbb P}$ and ${\mathbb Q}$ 
let us examine the denominator of the conditional probability obtained in 
(\ref{eq:5}), and define it to be 
\begin{eqnarray}
\Phi_t = \sum_j p_j 
\exp\left( \sigma x_j \xi_t-\half \sigma^2 x_j^2 t\right) . 
\end{eqnarray} 
Then an application of Ito's formula shows that 
\begin{eqnarray}
\frac{\rd\Phi_t}{\Phi_t} = \sigma {\hat X}_t \rd \xi_t \, ,
\end{eqnarray}
from which, upon integration and making use of the initial condition $\Phi_0=1$, 
it follows that 
\begin{eqnarray}
\Phi_t = \exp\left(\sigma\int_0^t {\hat X}_s \rd \xi_s - \half \sigma^2
\int_0^t {\hat X}_s^2\rd s\right). \label{eq:11.4}
\end{eqnarray}
Therefore, on account of Girsanov's theorem there exists an equivalent
probability measure ${\mathbb Q}$ over any finite time horizon such that 
the process $\{\xi_t\}$ defined by (\ref{eq:12}) is a standard Brownian motion 
in the ${\mathbb Q}$-measure, where $\{\Phi_t\}$ is the change-of-measure 
density martingale. In particular, for any measurable random variable $Y_t$ 
the conditional expectations in these two probability measures are related
according to 
\begin{eqnarray}
{\mathbb E}_s^{\mathbb P}[Y_t]=\frac{1}{\Phi_s}{\mathbb
E}_s^{\mathbb Q} [\Phi_t Y_t] \quad {\rm and}\quad {\mathbb
E}_s^{\mathbb Q}[Y_t] = \Phi_s{\mathbb E}_s^{\mathbb P}
\left[\frac{1}{\Phi_t} Y_t\right]. 
\label{eq:11.5}
\end{eqnarray}

\section{Probability of winning an election} 
\label{sec:4}

We have derived in (\ref{eq:5}) the dynamical process for the \textit{a posteriori} 
probability that the preferred candidate is the $k$-th one. With this at hand we 
can ask a range of 
quantitative questions, for instance, given that the upcoming election is in $T$ 
years time, what is the probability that candidate $k$ will secure more than 
$K$~\% of the votes. We now address this question in the simple 
case where there are only two dominant candidates. It is worth remarking that 
such a question cannot be answered without a dynamical model at hand. 

For a two-candidate election (or, equivalently, for a `yes-no' referendum), we may 
label the candidates using the binary system by calling them 0 and 1. In other 
words, we let $x_0=0$ and $x_1=1$; and accordingly, for the \textit{a priori} 
probabilities we set $p_0=p$ and $p_1=1-p$. Then the conditional expectation of 
the random variable $X$ is given by 
\begin{eqnarray}
{\hat X}_t = \sum_{i=0}^1 x_i {\mathbb P}(X=x_i|\xi_t)  = 
\frac{(1-p)\exp\left( \sigma \xi_t-
\frac{1}{2} \sigma^2 t\right)} {p + (1-p) 
\exp\left( \sigma \xi_t-\frac{1}{2} \sigma^2 t\right)} ,
\end{eqnarray} 
which has the interpretation of representing the \textit{a posteriori} expectation 
of the percentage share of the votes for candidate 1, because in the binary case 
we have ${\hat X}_t=\pi_{1t}$. 

We now examine the \textit{a priori} probability ${\mathbb P}({\hat X}_T>K)$ that 
candidate 1 will get more than $K~\%$ of the votes in the election to take place 
at time $T$ in the future. We shall make use of the fact that 
\begin{eqnarray}
{\mathbb P}\left({\hat X}_T>K\right) = {\mathbb E}\left[ {\mathds 1}\{ 
{\hat X}_T>K\} \right], 
\label{eq:20}
\end{eqnarray}
that is, the probability of an event can be calculated by taking the expectation of 
the indicator function for that event. To calculate the expectation in the right side 
of (\ref{eq:20}) we shall change the measure ${\mathbb P} \to {\mathbb Q}$ by 
use of the density martingale 
\begin{eqnarray}
\Phi_t =   p + (1-p) \exp\left( \sigma \xi_t-\half \sigma^2 t\right) . 
\end{eqnarray} 
Then we have 
\begin{eqnarray}
{\mathbb P}\left({\hat X}_T>K\right) = {\mathbb E}^{\mathbb Q}\left[ \Phi_T \, 
{\mathds 1}\{ {\hat X}_T>K\} \right] . 
\end{eqnarray}
We now observe that ${\hat X}_T$ is an increasing function of $\xi_t$. Thus the 
condition ${\hat X}_T>K$ is equivalent to a condition on $\xi_T$. This can be 
worked out explicitly. We have 
\begin{eqnarray}
(1-p)\exp\left( \sigma \xi_T-\half \sigma^2 T\right) > K \left[ 
p + (1-p)  \exp\left( \sigma \xi_T-\half \sigma^2 T\right) \right] , 
\end{eqnarray} 
from which it follows that ${\hat X}_T>K$ holds if and only if $\xi_T>z^*\sqrt{T}$, 
where 
\begin{eqnarray}
z^* = \frac{ \log\left( \frac{pK}{(1-p)(1-K)}\right) + \frac{1}{2}\sigma^2 T}
{\sigma \sqrt{T}} . 
\label{eq:24} 
\end{eqnarray} 

\begin{figure}[t!]
\centerline{
\includegraphics[width=0.75\textwidth]{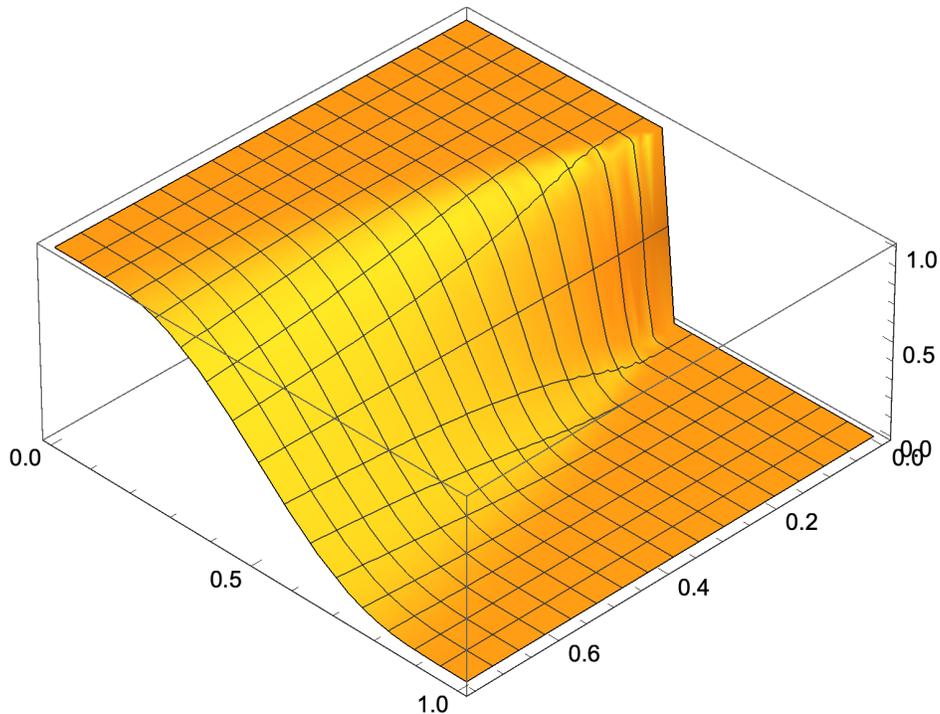}
}
\caption{\footnotesize{
\textit{Winning likelihood}. 
The probability that candidate 1 will win the election in a year ($T=1$), as a 
function of the \textit{a priori} probability $p\in(0,1)$ and the information flow 
rate $\sigma\in(0,0.75)$. In the limit $\sigma\to0$ where the future uncertainty 
is large, the probability approaches a step function, whereas in the opposite 
limit $\sigma\to\infty$ the probability approaches a linear function of the current 
poll represented by $p$. 
}}
\label{fig1}
\end{figure}

As indicated above, under the measure ${\mathbb Q}$ the information process 
$\{\xi_t\}$ is a standard Brownian motion, so we deduce that  
\begin{eqnarray} 
{\mathbb P}\left({\hat X}_T>K\right) = \frac{1}{\sqrt{2\pi}} \int_{z^*}^\infty 
\re^{-\frac{1}{2}z^2} \left( p + (1-p) \re^{\sigma \sqrt{T} z -\frac{1}{2} \sigma^2 T} 
\right) \rd z .
\end{eqnarray} 
Therefore, if we define $d^-=-z^*$ and $d^+=\sigma\sqrt{T}-z^*$, or more 
explicitly, 
\begin{eqnarray}
d^\pm =  \frac{ \log\left( \frac{(1-p)(1-K)}{pK}\right) \pm \frac{1}{2}\sigma^2 T}
{\sigma \sqrt{T}} ,
\label{eq:26} 
\end{eqnarray} 
then we have 
\begin{eqnarray} 
{\mathbb P}\left({\hat X}_T>K\right) = p\, N(d^-) + (1-p)\, N(d^+) , 
\label{eq:13}
\end{eqnarray} 
where 
\begin{eqnarray} 
N(x) =  \frac{1}{\sqrt{2\pi}} \int_{-\infty}^{x}  \re^{-\frac{1}{2}z^2} \rd z  
\end{eqnarray} 
denotes the cumulative normal distribution function. By setting $K=1/2$ in 
(\ref{eq:13}) we thus obtain the probability that candidate 1 will win the election. 
This is sketched in Figure~\ref{fig1}. We remark, incidentally, that 
the formula (\ref{eq:13}) thus obtained for the probability that a given candidate 
winning the election is essentially identical to the pricing formula in financial 
markets for an option on a stock in the Black-Scholes model \cite{BS}. 

\begin{figure}[t!]
\centerline{
\includegraphics[width=0.75\textwidth]{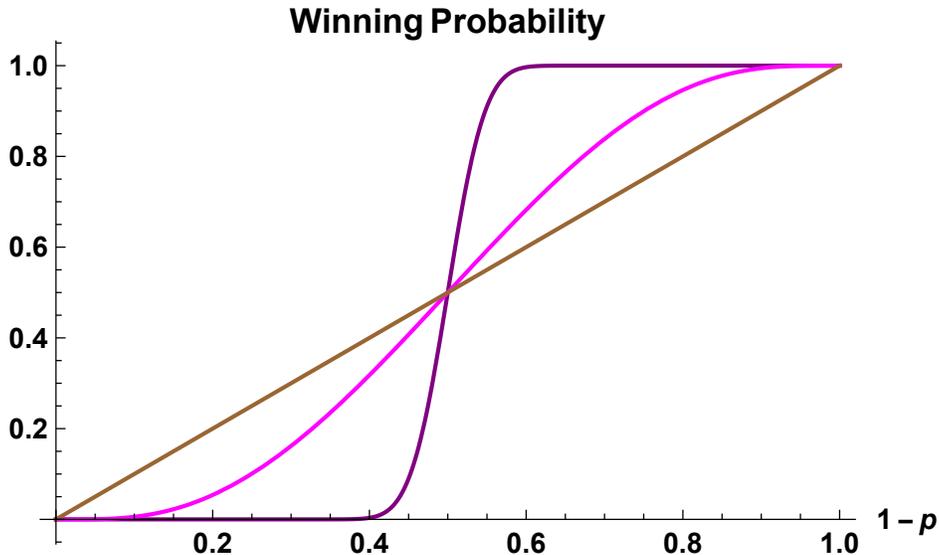}
}
\caption{\footnotesize{
\textit{Winning likelihood}. 
The probability that candidate 1 will win the election in a year ($T=1$), as a 
function of the current popularity rate $1-p$ for the candidate. If today's poll were 
the predictor for the winning probability, then the function should be a straight 
line shown here (in brown). However, according to the information-based model 
we see that the actual likelihood of winning the election is higher than what today's 
poll might suggest if $1-p>\frac{1}{2}$; and conversely lower than what the poll 
suggests if $1-p<\frac{1}{2}$. Two examples are shown, corresponding to the values 
$\sigma=0.15$ (in purple) and $\sigma=0.95$ (in magenta) for the information flow 
rate. The result shows that the deviation away from the poll indicator is greater 
if future uncertainty is greater (smaller $\sigma$). 
}}
\label{fig3}
\end{figure}

It is interesting to examine more closely the behaviour of the probability (\ref{eq:13}) 
of candidate 1 winning the future election. For instance, in the 
limit $\sigma\to0$ the probability as a function of the current poll approaches a step 
function. That is, if, say, candidate 1 has 51\% of support today, then in this limit the 
probability of candidate 1 winning the future election approaches one. This is because 
in the limit where the 
information flow rate going to zero, no information relevant to the election 
will be revealed. Hence, in the absence of any further information, the current state 
will be the future state, i.e. 51\% of the voters will vote for candidate 1, and hence the 
probability of winning approaches one. 

In reality, of course, information unravels, 
resulting in the dynamical evolution of the poll. In Figure~\ref{fig3} we plot the 
cross-section of (\ref{eq:13}) for two values of $\sigma$. If the current poll were 
the reflection of the election predictor, then the probability of a given candidate 
winning the future election would be a linear function of the current popularity, i.e. 
the current support rate equals the likelihood of election victory. However, according 
to the information-based model, the correspondence is nonlinear. In particular, if 
today's support rate of a candidate is greater than 50\%, then the likelihood of that 
candidate winning the future election is always greater than what is suggested by 
today's poll, and conversely if the current support rate is less than 50\% then the 
actual likelihood of winning is smaller than what is suggested by today's poll. 
Furthermore, the gap between today's poll and the winning likelihood increases as 
the future uncertainty increases.

\section{Predicting the election outcome} 
\label{sec:5}

In the foregoing analysis, we have introduced an abstract random variable $X$ 
that represents in some sense the `preferred' candidate. Consequently, the 
conditional expectation ${\hat X}_t$ of $X$ does not converge to any one of the 
candidates because the variability in public opinion remains wide leading up to 
the election day. From the viewpoint of a campaign manager, an election pollster, 
an election pundit, or a betting agency involved in elections, however, what 
matters is not so much about an abstract idea of which candidate might ultimately 
be judged by history to be the most ideal candidate. What matters to them is the 
more concrete notion of who might actually win the election. 

To model the prediction of an election we require a \textit{dynamical} version of 
the winning probability (\ref{eq:13}). To keep the discussion simple, let us for the 
moment continue on the assumption that there are only two candidates: 
candidate 0 and candidate 1. Then the probability that candidate 1, say, winning 
the election has to converge to either zero or one, depending on the election 
outcome. In other words, we need to consider the probability 
\begin{eqnarray}
{\mathbb P}_t\left({\hat X}_T>K\right) = {\mathbb E}_t\left[ {\mathds 1}\{ 
{\hat X}_T>K\} \right]
\label{eq:29}
\end{eqnarray}
\textit{conditional} on the information available up to time $t$. This conditional 
probability process will then evolve in a random manner such that it will converge 
to either zero or one, as the election day approaches, i.e. as $t\to T$. 

As a matter of clarification we remark that (\ref{eq:29}) represents a dynamical 
extension of the \textit{a priori} probability (\ref{eq:13}) in the sense that while 
(\ref{eq:13}) with $K=1/2$ represents the current (at time $t=0$) probability that 
candidate 1 will secure a win on the election day at $t=T$, this probability will 
change from day to day dynamically in accordance with the revelation of new 
information, so in particular given the information available at time $t$ the winning 
probability changes to (\ref{eq:29}) with $K=1/2$. 

To work out the conditional expectation in (\ref{eq:29}) we begin by remarking 
that the model under consideration entails a dynamic consistency in the following 
sense. Suppose that information has been gathered until time $s\in[0,t]$ so that 
the \textit{a priori} probability $p_i$ has changed into the \textit{a posteriori} 
probability $\pi_{is}={\mathbb P}(X=x_i|\xi_s)$. Then starting from this point, and 
given the knowledge of the past $\{\xi_u\}_{0\leq u\leq s}$, the reinitialised 
information process commencing from time $s$, according to the original model 
(\ref{eq:1}) for information flow, will take the form 
\begin{eqnarray}
\xi_{st} = \xi_t-\xi_s = \sigma X (t-s) + (B_t-B_s) . 
\end{eqnarray} 
Thus, starting from time $s$, what was the \textit{a posteriori} probability 
$\pi_{is}$ now 
becomes the \textit{a priori} probability for the future times $t\geq s$, so 
according to the logic leading to (\ref{eq:5}) the new \textit{a posteriori} 
probability $\pi_{it}={\mathbb P}(X=x_i|\xi_{t})$ should be of the form 
\begin{eqnarray}
\pi_{it} =\frac{\pi_{is}\exp\left( \sigma x_i \xi_{st}-
\frac{1}{2} \sigma^2 x_i^2 (t-s)\right)} {\sum_j \pi_{js} 
\exp\left( \sigma x_j \xi_{st}-\frac{1}{2} \sigma^2 x_j^2 (t-s)\right)} .
\label{eq:31}
\end{eqnarray}
Substituting the expressions for $\xi_{st}$ and $\pi_{is}$ in (\ref{eq:31}), a short 
calculation shows 
that the resulting expression indeed agrees with that obtained in (\ref{eq:5}), thus 
establishing dynamic consistency of the model. 

\begin{figure}[t!]
\centerline{
\includegraphics[width=0.75\textwidth]{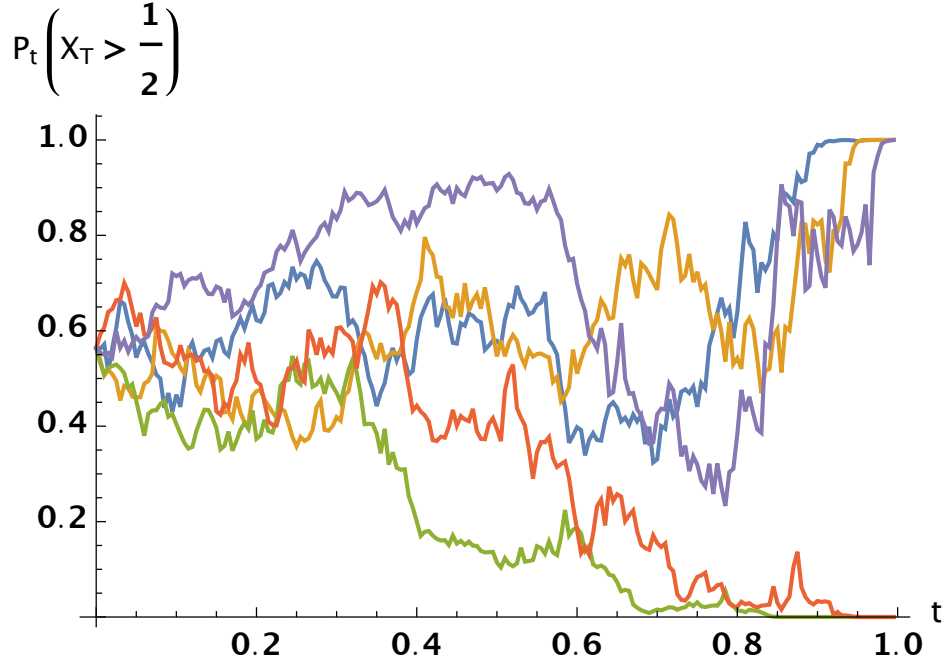}
}
\caption{\footnotesize{
\textit{Dynamics of election prediction}. 
The probability ${\mathbb P}_t\left({\hat X}_T>1/2\right)$ at time $t$ that 
candidate 1 will win the election to take place at time $T$ is simulated. Five 
sample paths are shown here, with the parameter values $p=0.48$, 
$\sigma=0.5$, and $T=1$. 
}}
\label{fig4}
\end{figure}

The dynamic consistency implies that to work out the conditional expectation in 
(\ref{eq:29}) it suffices to recycle the calculation leading up to (\ref{eq:13}) instead 
of performing a direct calculation making use of the time-dependent version of the 
measure change rule 
${\mathbb E}_t^{\mathbb P}[Y_T]={\mathbb E}_t^{\mathbb Q} [\Phi_T Y_T]/\Phi_t$. 
In particular, we deduce at once that 
\begin{eqnarray} 
{\mathbb P}_t\left({\hat X}_T>K\right) = \pi_t\, N(d_t^-) + (1-\pi_t)\, N(d_t^+) , 
\label{eq:32}
\end{eqnarray} 
where $\pi_t={\mathbb P}(X=0|\xi_{t})$ and 
\begin{eqnarray}
d_t^\pm =  \frac{ \log\left( \frac{(1-\pi_t)(1-K)}{\pi_t K}\right) \pm 
\frac{1}{2}\sigma^2 (T-t)}{\sigma \sqrt{T-t}} . 
\end{eqnarray} 
Setting $K=1/2$ in (\ref{eq:32}) we obtain the \textit{a posteriori} probability 
that candidate 1 will win the election to take place at time $T$, and this indeed 
converges to either zero or one, depending on how information unravels along 
the way. Figure~\ref{fig4} shows five sample-path simulations of the conditional 
probability process. We remark, incidentally, that if there are more than two 
candidates, say, $N$ candidates competing, then the corresponding 
\textit{a posteriori} probability at time $t$ that the $k$th candidate winning the 
election to take place at time $T$ is determined by computing 
${\mathbb P}_t(\pi_{kT}>N^{-1})$. 

\section{Impact of disinformation: when to release fake news?}  
\label{sec:6}

The information-based model for election naturally allows for a generalisation 
to model situations in which there are disinformation in circulation, with a 
deliberate intent to obscure the true value of $X$. In particular, in \cite{BM} we 
have defined what one might mean by `fake news' as a modification of the 
information process (\ref{eq:1}) in the following form: 
\begin{eqnarray} 
\eta_t = \sigma X t + B_t + F_t, 
\end{eqnarray} 
where the term $\{F_t\}$, which has a bias so that ${\mathbb E}[F_t]\neq0$, 
models deliberate disinformation. The idea can be described as follows. There 
are unfounded rumours 
and speculations obscuring the value of $X$, but a large number of such 
random speculations will average over to give rise to an unbiased noise so that 
${\mathbb E}[B_t]=0$. In other words, while noise will interfere with giving an 
accurate estimate for $X$, it is not directed in any particular orientation; whereas 
fake news can be distinguished from conventional noise 
by its desire to disorient the public. Thus, those 
who are not aware of the 
existence of $\{F_t\}$ will arrive at their estimates based on formula 
(\ref{eq:5}), but with the distorted information 
$\{\eta_t\}$ in place of $\{\xi_t\}$. In other words, they 
will `perceive' the information as taking the normal form (\ref{eq:1}) and proceed 
to make appropriate inferences based on (\ref{eq:5}); but their inferences are 
now skewed owing to the existence of $\{F_t\}$. 

In the information-based model, the effect of disinformation can be understood 
in an intuitive and transparent manner: if ${\mathbb E}[F_t]>0$ then people 
(unaware of the 
existence of $\{F_t\}$) are misguided to thinking that the true value of $X$ is 
greater than what it really is; and similarly, if ${\mathbb E}[F_t]<0$ then people 
are misguided to thinking that the true value of $X$ is less than what it really is. 
By choosing specific models for the process $\{F_t\}$ one can therefore apply 
a simulation study to determine, for a given choice of $\{F_t\}$, how the 
opinion-poll statistics might be affected in that situation. 

From the viewpoint of those who wish to disseminate disinformation, the most 
obvious question that arises is: How to find optimum choice for $\{F_t\}$? 
Evidently, the notion of optimality depends on the choice of the criteria, but in 
the present context perhaps the most natural one is that maximises 
the probability of a given candidate winning the election. In general, finding a 
solution to this question requires solving a new type stochastic optimisation 
problem that combines both (a) the theory of signal detection and in particular 
nonlinear filtering \cite{kai}, and (b) the theory of change-point detection problem 
\cite{Shiryaev}. Thus, we encounter a situation here in which a new type of 
problem in society leads to a new type of mathematical challenge. 

Here we shall analyse this problem in a simple setup in which there is only one 
chance for disseminating fake news. Hence the problem is to find the optimal 
timing to release fake news so as to maximise its impact on the upcoming 
election. We shall assume, in particular, a model for fake news of the form 
$F_t = \mu (t-\tau) \, \re^{-\alpha(T-\tau)}\, {\mathds 1}\{t>\tau\}$, where $\mu$ 
and $\alpha>0$ are constants, $\tau$ denotes the 
time at which fake news are released, and ${\mathds 1}\{t>\tau\}$ as before 
denotes the indicator function so that ${\mathds 1}\{t>\tau\}=0$ if $t\leq\tau$ and 
${\mathds 1}\{t>\tau\}=1$ otherwise. This choice of $F_t$ has the interpretation 
that when fake news are released at time $\tau$, initially their strengths grow 
linearly in time at the rate $\mu$, but over time the strengths of fake news get 
suppressed exponentially at the rate $\alpha$. 

Let us consider the problem in the context of a two-candidate election. Recall 
that in the absence of disinformation the probability of candidate 1 
securing an election win has been worked out in (\ref{eq:13}) with $K=1/2$. If, 
however, disinformation is disseminated in such a way that the voters are 
unaware of this, then this probability is altered in the following way. Noting that 
(\ref{eq:13}) has been obtained by assuming a genuine information flow 
$\{\xi_t\}$, if in reality this is replaced by $\{\eta_t\}$, then the threshold value 
$z^*$ of (\ref{eq:24}) is now replaced by $z^*-F_T/\sqrt{T}$. This follows on 
account of the fact that the original condition was $\xi_T>z^*\sqrt{T}$, but 
in the presence of disinformation $\xi_T$ is replaced by $\xi_T+F_T$; hence 
the condition now reads $\xi_T>(z^*-F_T/\sqrt{T})\sqrt{T}$. Accordingly, 
the variables $d^\pm$ are replaced by $d^\pm+F_T/\sqrt{T}$. The effect of 
this is to increase (respectively, decrease) the winning probability 
${\mathbb P}({\hat X}_T>1/2)$ for candidate 1 if $F_T>0$ (respectively, if 
$F_T<0$). Thus, 
in principle one can optimise the form of $\{F_t\}$ so as to maximise (or 
minimise) the probability ${\mathbb P}({\hat X}_T>1/2)$. However, in reality, 
owing to the Markovianity of $\{\xi_t\}$ along with the assumption that the 
existence of $\{F_t\}$ is hidden from the voters, maximum impact is obtained 
in this case by simply maximising $F_T$. In the case of the model 
$F_t=\mu(t-\tau)\, \re^{-\alpha(T-\tau)}\, {\mathds 1}\{t>\tau\}$, therefore, to 
achieve a maximum impact on the election outcome, fake news should be 
released at 
\begin{eqnarray}
\tau^* = \frac{\alpha T-1}{\alpha} 
\end{eqnarray}
if $\alpha>T^{-1}$, and at $\tau^*=0$ if $\alpha\leq T^{-1}$. 

\begin{figure}[t!]
\centerline{
\includegraphics[width=0.75\textwidth]{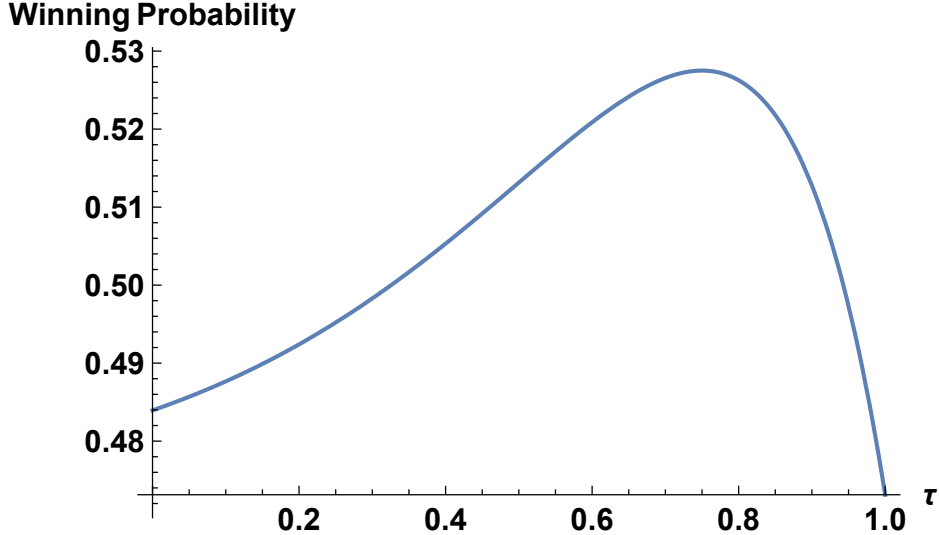}
}
\caption{\footnotesize{
\textit{Winning likelihood}. 
The probability that candidate 1 will win the election to take 
place in one year ($T=1$), as a function of 
the timing $\tau$ at which fake news in favour of candidate 1 are released, is plotted, 
when we set $\alpha=5$. The figure shows that by releasing fake news four-fifth 
of the way into the election period, the probability of candidate 1 winning the 
election can be enhanced by about 5.5\% (from about 47.3\% to about 52.8\%). 
Other model parameters are: 
$\sigma=0.3$ for the information flow rate, $p=0.505$ for the \textit{a priori} 
probability that candidate 0 wins the election, and $\mu=1.5$ for the fake news 
strength. 
}}
\label{fig2}
\end{figure}

In Figure~\ref{fig2} 
we plot the probability ${\mathbb P}({\hat X}_T>1/2)$ of candidate 1 winning 
the election in the presence of fake news as a function of the release timing 
$\tau$. For the parameter choices made therein, we see that in the absence of 
fake news, if the \textit{a priori} probability (today's opinion poll) for candidate 
1 to win the election is $1-p=49.5$\%, then the actual probability (today's prediction) 
of winning the election 
one year later is about 47.3\%. However, the release of fake news in favour 
of candidate 1, if undetected by the voters, will enhance this likelihood 
whenever it is released prior to the election day. If, in particular, the release 
timing is optimised, then 
the probability can be enhanced by as much as 5.5\% (in this example), perhaps 
just sufficient to overcome statistical uncertainties for candidate 1 to secure a 
victory. The specific figures mentioned here are of course based on an arbitrarily 
chosen model parameters, but the model clearly illustrates the impact of fake 
news in an intuitive manner, and allows for a more comprehensive impact 
studies, scenario analysis, as well as analysis on parameter sensitivity; the 
results of which we hope will then be fed into the development of counter 
measures to tackle the impact of fake news.

\section{Managing information flow in an election campaign} 
\label{sec:7} 

In the previous section, and more generally in \cite{BM}, we examined the impact 
of the dissemination of disinformation on the dynamics of the opinion poll statistics. 
It should be apparent, however, that the framework is not restricted to modelling 
disinformation. For instance, if the campaign team is confident about the value of 
$X$, then they can proactively promote (or perhaps to hide) relevant information, 
but not in secret. 
In the simplest situation one could set $F_t=\kappa X (t-\tau) {\mathds 1}\{t>\tau\}$ 
for some $\kappa>0$, where $\{F_t\}$ now represents genuine information. 
Then from time $\tau$ onwards the voters will obtain more reliable information 
about $X$ than otherwise. This is equivalent to having a time-dependent 
information flow rate $\sigma_t$ such that $\sigma_t=\sigma$ for $t\leq\tau$ and 
$\sigma_t=\sigma+\kappa$ for $t>\tau$. More generally, we may consider a generic 
time-dependent information flow rate $\sigma_t$. If the campaign team is in the 
position to control certain information, then they would naturally like to optimise the 
way in which information is managed so as to maximise the chances of their 
candidate winning 
the election. (In fact, as explained in the final section below, such a situation is very 
natural within the structural approach; whereas in the reduced-form approach the 
random variable $X$ is to an extent an abstraction so it is not always apparent 
whether the campaign team can manage information on $X$. However, because 
the mathematical treatment of the problem in either of the frameworks 
is the same, and because the present 
paper is concerned with the reduced-form model, we shall proceed to develop the 
idea here, with the caveat that practical implementations of the ideas presented in 
this section are more suitable within the structural framework.) 

In the general case for which the information flow rate is time dependent, the 
information process takes the form 
\begin{eqnarray}
\xi_t = X \int_0^t \sigma_s \rd s + B_t . 
\end{eqnarray} 
In this case, the information process $\{\xi_t\}$ is no longer Markovian so that we 
do not have the simplifying reduction ${\mathbb P}(X=x_i|\{\xi_s\}_{0\leq s\leq t}) 
\to {\mathbb P}(X=x_i|\xi_t)$. In other words, the conditional probability for the 
random variable $X$ is now path dependent. To work out the posterior probabilities 
we begin by remarking that if we define the process $\{\Phi_t\}$ according to 
\begin{eqnarray}
\Phi_t = \exp\left( X\int_0^t \sigma_s \rd\xi_s - \frac{1}{2}\int_0^t\sigma_s^2\rd s 
\right) ,
\end{eqnarray} 
then we can use the unit-initialised martingale $\{\Phi_t\}$ to change probability 
measure ${\mathbb P}\to{\mathbb Q}$ with the property that (i) the information 
process $\{\xi_t\}$ is a Brownian motion under ${\mathbb Q}$ independent of 
$X$; (ii) the random variable $X$ has the same probability law under 
${\mathbb Q}$ as it does under ${\mathbb P}$; and (iii) the conditional 
expectation $f_t={\mathbb E}^{\mathbb P}[f(X)|\{\xi_s\}_{0\leq s\leq t}]$ for a 
function of the random variable $X$ can be obtained by use of the 
Kallianpur-Striebel \cite{KS} formula  
\begin{eqnarray}
f_t = \frac{{\mathbb E}^{\mathbb Q}[f(X) \Phi_t|\{\xi_s\}_{0\leq s\leq t}]}
{{\mathbb E}^{\mathbb Q}[\Phi_t|\{\xi_s\}_{0\leq s\leq t}]} . 
\label{eq:38} 
\end{eqnarray}
By setting $f(X)={\mathds 1}\{X=x_i\}$ in (\ref{eq:38}) we thus deduce the 
expression for the conditional probability $\pi_{it}=
{\mathbb P}(X=x_i|\{\xi_s\}_{0\leq s\leq t})$ with time-dependent information 
flow: 
\begin{eqnarray}
\pi_{it} = \frac{p_i \exp\left( 
x_i \int_0^t \sigma_s {\rm d}\xi_s - \frac{1}{2}x_i^2\int_0^t \sigma_s^2 
{\rm d}s\right)}{\sum_j p_j \exp\left( 
x_j \int_0^t \sigma_s {\rm d}\xi_s - \frac{1}{2}x_j^2\int_0^t \sigma_s^2 
{\rm d}s\right)} ,
\end{eqnarray} 
from which the best estimate ${\hat X}_t$ for $X$ can be obtained according 
to ${\hat X}_t = \sum_i x_i \pi_{it}$. 

Let us proceed further by considering, as in Section~\ref{sec:4}, the case in 
which there are only two candidates. Then we have 
\begin{eqnarray}
{\hat X}_T =
\frac{(1-p) \exp\left( 
\int_0^T \sigma_s {\rm d}\xi_s - \frac{1}{2}\int_0^T \sigma_s^2 
{\rm d}s\right)} {p + (1-p) 
\exp\left( 
\int_0^T \sigma_s {\rm d}\xi_s - \frac{1}{2}\int_0^T \sigma_s^2 
{\rm d}s\right)} ,
\label{eq:40}
\end{eqnarray} 
which is an increasing function of a single Gaussian random variable 
$\int_0^T \sigma_s {\rm d}\xi_s$. It follows that ${\hat X}_T>K$ holds if and 
only if 
\begin{eqnarray}
\frac{\int_0^T \sigma_s {\rm d}\xi_s}{\sqrt{\int_0^T \sigma_s^2 
{\rm d}s}} > 
\frac{ \log\left( \frac{pK}{(1-p)(1-K)}\right) + \frac{1}{2}\int_0^T \sigma_s^2 
{\rm d}s}
{\sqrt{\int_0^T \sigma_s^2 
{\rm d}s}} . 
\end{eqnarray} 
On account of the relation ${\mathbb P}({\hat X}_T>K)=
{\mathbb E}^{\mathbb Q}[\Phi_T {\mathds 1}\{{\hat X}_T>K\}]$, where 
$\{\Phi_T\}$ is the denominator of (\ref{eq:40}), 
a short calculation then shows that 
\begin{eqnarray} 
{\mathbb P}\left({\hat X}_T>K\right) = p\, N(d^-) + (1-p)\, N(d^+) , 
\label{eq:42} 
\end{eqnarray} 
where 
\begin{eqnarray}
d^\pm =  \frac{ \log\left( \frac{(1-p)(1-K)}{pK}\right) \pm \frac{1}{2}\int_0^T 
\sigma_s^2 {\rm d}s}{\sqrt{\int_0^T \sigma_s^2 {\rm d}s}} . 
\end{eqnarray} 
Setting $K=1/2$ in (\ref{eq:42}) we thus obtain the probability that candidate 1 
wins the future election, when information flow rate $\{\sigma_t\}$ is time 
dependent. 

From the viewpoint of the campaign team for candidate 1, if they had the ability 
to control the rate of information flow, then it would be desirable to find a 
$\{\sigma_t\}$ that maximises the success probability (\ref{eq:42}). The basic 
idea can already be inferred from inspecting Figure~\ref{fig3} in the case of 
constant $\sigma$: depending on the value of the \textit{a priori} probability $p$, 
one would like to either let $\sigma\to\infty$ or to let $\sigma\to0$ so as to impact 
the probability of a given candidate winning the future election. These 
extreme cases, however, are unrealistic (even in the structural model): release 
of information (marketing) is in general costly. It is therefore reasonable to 
assume that, if the campaign period is $[0,T]$, then the number 
$\int_0^T \sigma_s\, {\rm d}s$ is strictly bounded. 

With this in mind, we examine how the winning probability $P=
{\mathbb P}({\hat X}_T>1/2)$ for candidate 1 changes as the information flow 
rate $\sigma_t$ 
is varied. To this end let us consider the functional derivative of $P$ with respect 
to $\sigma_t$. That is, regarding $P=P[\sigma]$ as a functional of $\sigma$, 
we consider perturbing $\sigma_t$ by a small amount 
$\epsilon$ in the direction of $\phi_t$:
\begin{eqnarray}
\frac{\delta P}{\delta\sigma} = \lim_{\epsilon\to0}  
\frac{P[\sigma+\epsilon \phi]-P[\sigma]}{\epsilon} . 
\end{eqnarray} 
Then a short calculation shows that 
\begin{eqnarray} 
\frac{\delta P}{\delta\sigma} =\frac{1}{\sqrt{2\pi}} \left[ p\, \re^{-\frac{1}{2}(d^-)^2} 
\, \frac{\delta d^-}{\delta\sigma} + (1-p)\, \re^{-\frac{1}{2}(d^+)^2} \, 
\frac{\delta d^+}{\delta\sigma} \right], 
\label{eq:45} 
\end{eqnarray} 
where 
\begin{eqnarray}
\frac{\delta d^\pm}{\delta\sigma} = 
\frac{\int_0^T \sigma_s \phi_s {\rm d}s}{\sqrt{\int_0^T\sigma_s^2{\rm d}s}} 
\left[ \frac{\log\left( \frac{pK}{(1-p)(1-K)}\right)}{\int_0^T \sigma_s^2{\rm d}s} 
\pm 1\right] .
\end{eqnarray}
From (\ref{eq:45}) we deduce that the perturbation on $P$ is proportional to the 
inner product $\int_0^T \sigma_s \phi_s {\rm d}s$ between $\{\sigma_t\}$ and 
$\{\phi_t\}$. Therefore, for a given $\{\sigma_t\}$ one can explore how it may be 
perturbed so as to either increase or decrease $P$. In practical applications, 
however, 
it is likely that one would be working with a parametric family of models for the 
information flow rates $\{\sigma_t\}$, and in this case optimisation can be achieved 
with normal differentiation, not with a functional derivative. 

We note, more generally, that the way in which the winning probability $P$ 
changes against a perturbation of the information flow rate today will not be the 
same as what it is tomorrow. Indeed, the \textit{a posteriori} probability 
$\pi_t={\mathbb P}(X=x_0|\{\xi_s\}_{0\leq s\leq t})$ 
tomorrow may change from the \textit{a priori} probability $p$ today in such a way 
that while there is an advantage in increasing $\sigma$ today, it would be 
advantageous to decrease $\sigma$ tomorrow. In other words, what we require 
is a dynamical version of (\ref{eq:45}) based on the conditional version of the 
winning probability: $P_t={\mathbb P}({\hat X}_T>1/2|\{\xi_s\}_{0\leq s\leq t})$. 
This can be worked out straightforwardly, and we deduce that the result takes a 
form identical to (\ref{eq:45}), except that $p$ is replaced by $\pi_t$, 
$\int_0^T \sigma_s^2{\rm d}s$ is replaced by $\int_t^T \sigma_s^2{\rm d}s$, and 
$\int_0^T \sigma_s \phi_s {\rm d}s$ is replaced by $\int_t^T \sigma_s \phi_s {\rm d}s$.

\section{Sensitivity analysis} 
\label{sec:8} 

Another aspect of the information-based model that will be useful to explore for 
strategic planning is concerned with parameter sensitivity. If, for instance, 
releasing of information is costly (e.g., advertising cost), and if the result is not 
likely to significantly change the state of affairs, then it may not be advantageous 
to proceed with the release. For such a consideration, knowledge of the parameter 
sensitivity of the model will help in assisting decision making. 

\begin{figure}[t!]
\centerline{
\includegraphics[width=0.75\textwidth]{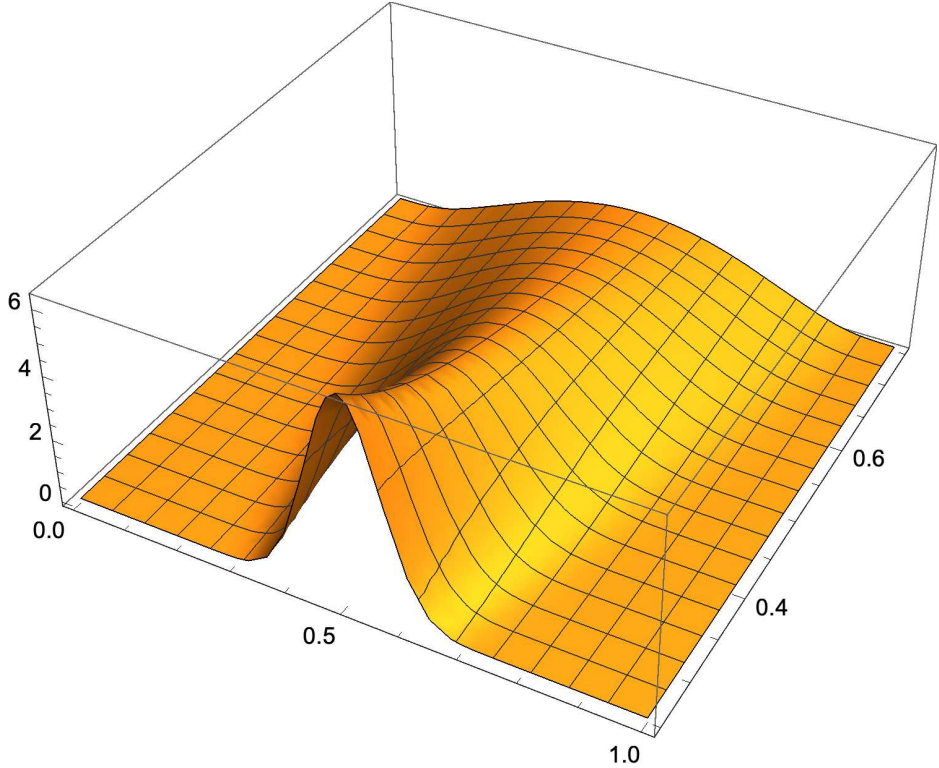}
}
\caption{\footnotesize{
\textit{Density function for ${\hat X}_T$}. 
The \textit{a priori} probability density $\rho(x)$ for the random variable 
${\hat X}_T$ 
is shown here for different values of the information flow-rate parameter 
$\sigma\in[0.25,0.75]$. Other model parameters are $p=0.5$ and $T=1$. 
}}
\label{fig5}
\end{figure}

In the present context, we are interested, in particular, in the parameter sensitivity 
of the probability distribution of the future random variable ${\hat X}_T$. Let us 
therefore begin by working out the \textit{a priori} density function for ${\hat X}_T$. 
Recalling that the density function $\rho(x)$ for ${\hat X}_T$ is given by $\rho(x)=
\rd {\mathbb P}({\hat X}_T<x)/\rd x$, we deduce, by differentiating (\ref{eq:13}), 
that 
\begin{eqnarray}
\rho(x) = \frac{1}{\sqrt{2\pi\sigma^2T} x(1-x)} \left[ p\,\re^{-\frac{1}{2}(d^-)^2} 
+ (1-p) \re^{-\frac{1}{2}(d^+)^2} \right] ,
\label{eq:47} 
\end{eqnarray}
where 
\begin{eqnarray}
(d^\pm)^2 = \left( 
\frac{ \log\left( \frac{(1-p)(1-x)}{px}\right)}{\sigma \sqrt{T}} \right)^2 + \frac{1}{4} 
\sigma^2 T \pm \log\left( \frac{(1-p)(1-x)}{px}\right)  .
\end{eqnarray}
In Figure~\ref{fig5} we plot the density function $\rho(x)$ for different values of 
the parameter $\sigma$. It may not be immediately apparent from (\ref{eq:47}) 
that $\rho(x)$ is a density function satisfying the normalisation condition 
$\int_0^1\rho(x)\rd x=1$ (although by definition $\rho(x)$ clearly is a density 
function), but in fact $\rho(x)$ is a Gaussian mixture density in a transformed 
variable. To see this, let use change the variable $x\to u$ according to 
\begin{eqnarray} 
u = \log\left( \frac{1-x}{x}\right) , 
\end{eqnarray} 
and likewise to simplify the notation let us define $\beta=\log[(1-p)/p]$. Then it 
should be immediately apparent that while $x$ varies from $0$ to $1$, $u$ 
varies from $\infty$ to $-\infty$. Together with the fact that $\rd u = -\rd x /x(1-x)$ 
we thus deduce that the density function $f(u)$ for the transformed variable 
$U_T=\log[(1-{\hat X}_T)/{\hat X}_T]$ takes the form 
\begin{eqnarray} 
f(u) = \frac{1}{1+\re^\beta} \frac{1}{\sqrt{2\pi\sigma^2T}} \left[ 
\re^{-\frac{1}{2\sigma^2T} \left( u - (\beta - \sigma^2 T/2)\right)^2} + 
\re^\beta 
\re^{-\frac{1}{2\sigma^2T} \left( u - (\beta + \sigma^2 T/2)\right)^2} \right] ,
\label{eq:50}
\end{eqnarray} 
which we recognise at once to be a normalised Gaussian mixture density. 

With the expression for the density function at hand we are in the position 
to address the question regarding parameter sensitivity. We are interested, 
in particular, in the sensitivity of the model against the information flow rate 
$\sigma$ and the current poll statistics $p$, for, $\sigma$ can be directly 
linked to the advertisement cost, while the value of $p$ can be deduced 
from pollsters. For this analysis we shall borrow ideas from information 
geometry (see, e.g., \cite{Rao,DBDH}) to determine the parameter sensitivity, 
i.e. we shall examine the Fisher information matrix (or the Fisher-Rao metric) 
associated with the parameters $\sigma$ and $p$. The Fisher-Rao 
information metric $G_{ij}(\sigma,p)$ is useful inasmuch as it 
introduces the notion of a metric in the space of density functions that 
determines the relative separation of the densities for different parameter 
values. In other words, the amount of impact caused by changing, say, 
the value of the information revelation rate from $\sigma$ to $\sigma'$ is 
in general not related to the naive difference $\sigma-\sigma'$. Instead, it 
is measured in terms of the geodesic distance associated with the Fisher-Rao 
information metric. Writing $\theta^1=\sigma$ and $\theta^2=p$, the 
Fisher information is determined by expression 
\begin{eqnarray}
G_{ij}(\sigma,p) = \int_{-\infty}^\infty \frac{1}{f(u)} 
\frac{\partial f(u)}{\partial \theta^i}\frac{\partial f(u)}{\partial \theta^j} 
\rd u .
\label{eq:51}
\end{eqnarray} 
It should be noted here that the Fisher-Rao metric, unlike invariant 
quantities such as the curvature, is not strictly invariant under the variable transformation $X_T\to U_T$. However, under certain regularity conditions, 
Chentsov's theorem shows that the Fisher information is, up 
to scaling, the only invariant metric under certain statistically important 
transformations \cite{chentsov}. Hence, without loss but for the gain of computational simplicity we shall investigate the 
Fisher information associated with the density function $f(u)$, rather than that 
for $\rho(x)$. Specifically, if we substitute (\ref{eq:50}) in (\ref{eq:51}), then 
after a short calculation we deduce that 
\begin{eqnarray} 
G_{ij}(\sigma,p) = \left( \begin{array}{cc} 
T + \frac{2}{\sigma^2} & \frac{2p-1}{\sigma p(1-p)} \\ 
\frac{2p-1}{\sigma p(1-p)} & 
~~\frac{1}{p(1-p)}\left(1+ \frac{1}{\sigma^2Tp(1-p)} 
\right) \end{array} \right) . 
\end{eqnarray} 
On inspection of the one-dimensional subspace parameterised by the 
information revelation rate $\sigma$, the corresponding Fisher information 
is just a constant $T$ plus the the Fisher information $2/\sigma^2$ associated 
with the normal density function of standard deviation $\sigma$, and this is 
decreasing in $\sigma$, indicating the that the 
smaller the $\sigma$ is, the greater is the impact (associated with changing the 
value of $\sigma$) on the distribution of the projected future values of $X$. 
Putting the matter differently, if there is already a lot of reliable information 
being transmitted to the electorate, then spending campaign funding on further 
advertisements is probably not advisable because it will entail little additional 
impact; whereas if there is very little information being transmitted, then it is 
worth engaging in an additional advertisement campaign. This conclusion may 
be intuitively obvious, however, what is less obvious from intuition alone is where 
one draws the line between these two extremes. The advantage of working out the 
Fisher information is that it provides a quantitative measure that allows one to 
analyse such a problem. In particular, as regards the sensitivity of the 
distribution on the information revelation parameter, the geodesic distance between the densities associated with the parameter values $\sigma$ and 
$\sigma'$ can be worked out in closed form (see \cite{DBDH} ), which can be 
used to conduct a quantitative analysis on the cost-benefit analysis.

\section{Towards election planning: structural approach} 
\label{sec:9}

The two modelling approaches introduced in \cite{BM} are similar to the structural 
and reduced-form models used in credit-risk management in finance and 
investment banking context \cite{BR}. In the financial context, reduced-form models 
are commonly used 
because for a given financial contract, the number of cash flows and the number 
of independent market factors related to it are typically so vast that it is impractical 
to even attempt to dissect the product so as to identify its structural details. In 
contrast, in the reduced-form approach it is possible to capture abstractly 
all the qualitative 
features resulting from structural models, without going into any of the structural 
details, and this makes the reduced-form approach more practical to implement. 
In the context of election modelling, on 
the other hand, the structural approach is entirely feasible, for, (a) typically in an 
election there are only a small number of plausible candidates; and (b) the number 
of important independent issues relevant to a large number of voters is also 
relatively small. 

In the present paper we have focused our attention on reduced-form models 
because (i) it captures all the qualitative features arising from structural models, 
thus making it an ideal framework to conduct academic study of the system; (ii) 
one can pursue mathematical analysis quite far without cluttering it with all the 
structural details; and (iii) the mathematical formalisms and derived formulae 
carry through directly to 
structural models. However, if one were to apply the information-based 
framework as a part of election planning, then structural models are considerably 
more advantageous, for, there is a degree of abstraction in the reduced-form model 
that, in some situations, makes it difficult to implement in practice. For instance, 
in the 
discussion in Section~\ref{sec:7} on controlling the flow of information, this is 
feasible if the information is concerned specifically with the candidate's own 
positions on key issues, which would be the case in the structural approach. 
An analogous statement can be made on the material considered in 
Section~\ref{sec:8}. 
Thus, while the features of the information-based model investigated here 
for electoral 
competition can be well explored within the reduced form approach presented 
above, for a 
realistic implementation it is preferable to apply the techniques outlined in the 
present paper in the structural setup of \cite{BM}, 
where independent factors have 
realistic and tangible interpretations. 
In this connection it is worth adding that for model calibration one may 
use (${\alphaup}$) the current poll statistics to estimate the \textit{a priori} 
probabilities $\{p_i\}$; (${\betaup}$) the magnitude of the volatility for the 
poll statistics to estimate the information flow rate $\sigma$; and 
(${\gammaup}$) various public surveys conducted by pollsters to estimate 
the density for the preference weights $\{w_n^k\}$. In the presence of 
disinformation, the distribution of $\{F_t\}$ can be estimated from data 
available to fact checkers. 

With these in mind, we conclude by remarking that the analysis presented in 
this paper is in fact applicable more generally to generic advertisements; not 
merely to election competition. From this point of view we can regard the 
foregoing material as providing a new information-based mathematical model 
for controlling information, as well as for 
characterising the impact of advertisement.

\vspace{0.45cm} 
\noindent 
{\bf Acknowledgements}.
The author thanks Lane Hughston, David Meier and Bernhard Meister 
for discussion on related ideas, and anonymous referees for helpful 
suggestions.

\end{document}